\documentclass[aps,prb,preprint,superscriptaddress,longbibliography]{revtex4-1}

\usepackage{amsmath}
\usepackage{graphicx}
\usepackage{placeins}

\mathchardef\mhyphen="2D

\begin{document}

\title{Ultrafast Charge Transfer at a Quantum Dot/2D Materials Interface Probed by Second Harmonic Generation}

\author{Goodman, A.J.}
\affiliation{Department of Chemistry, Massachusetts Institute of Technology}
\author{Dahod, N.S.}
\affiliation{Department of Chemical Engineering, Massachusetts Institute of Technology}
\author{Tisdale, W.A.}
\affiliation{Department of Chemical Engineering, Massachusetts Institute of Technology}
\email[]{tisdale@mit.edu}

\date{\today}

\begin{abstract}
Hybrid quantum dot (QD) / transition metal dichalcogenide (TMD) heterostructures are attractive components of next generation optoelectronic devices, which take advantage of the spectral tunability of QDs and the charge and exciton transport properties of TMDs. Here, we demonstrate tunable electronic coupling between CdSe QDs and monolayer WS$_2$ using variable length alkanethiol ligands on the QD surface. Using femtosecond time-resolved second harmonic generation (SHG) microscopy, we show that electron transfer from photoexcited CdSe QDs to single-layer WS$_2$ occurs on ultrafast (50 fs - 1 ps) timescales. Moreover, in the samples exhibiting the fastest charge transfer rates ($\leq$50 fs) we observed oscillations in the time-domain signal corresponding to an acoustic phonon mode of the donor QD, which coherently modulates the SHG response of the underlying WS$_2$ layer. These results reveal surprisingly strong electronic coupling at the QD/TMD interface and demonstrate the usefulness of time-resolved SHG for exploring ultrafast electronic-vibrational dynamics in TMD heterostructures. 
\end{abstract}

\pacs{}
\maketitle

Hybrid structures containing both quantum dots (QDs) and 2D transition metal dichalcogenides (TMDs) leverage both the constituent 0D and 2D materials' favorable properties. 2D TMDs contribute high charge carrier mobilities\cite{MoS2_transistor,MoS2Transport,TMD_mobility_review} to mixed-dimensional heterostructures, while 0D QDs can provide strong, spectrally tunable broad-band absorption and efficient narrowband emission. Through careful selection of the 2D material and QD, the hybrid structure can be tuned to a specific application. Electronically insulated QDs have been used to sensitize 2D materials \textit{via} resonant energy transfer, revealing anomalous energy transfer phenomena at 0D/2D interfaces\cite{Graphene_FRET_UniversalSca,MoS2_FerryPaper_NanoLett2014,MoS2_BrusHEinz_Nanolett2016,Cotlet_EnergyTransfer} and enabling tunable interfacial coupling.\cite{MoS2_QD_electricallyControlledFRET_2015nanolett} Furthermore, photodetectors with large gain and facile spectral tunability have been demonstrated by strongly coupling QDs to 2D materials by exchanging the QD native ligands to short, conductive spacer ligands.\cite{GrapheneQD_phototrans,MoS2_PbS_photodetector} 

Recently, Boulesbaa \textit{et al.} reported observing ultrafast ($<45$ fs) charge transfer at the interface of monolayer WS$_2$ and CdSe QDs.\cite{0D2D_ChargeTransfer_JACS2016} This interpretation is surprising because the CdSe cores used in the study were passivated by a 2 nm thick insulating ZnS shell and 18-carbon octadecylamine ligands ($\sim2$ nm dot-to-dot spacing). Ultrafast charge transfer requires wave function overlap between donor and acceptor, and the presence of thick shells and long-chain ligands normally prevents strong electronic coupling in QDs.\cite{ColloidQD_prospects_Review} We note that QD/2D hybrid interfaces are difficult to probe with traditional linear spectroscopies, since both the QD and TMD possess large absorption cross sections with overlapping spectral features.\cite{cotlet_adv_fun_mater} In addition, 2D materials are extremely sensitive to the surrounding dielectric environment,\cite{MoS2_DielectricScreening_NanoLett2014,WS2_nonhydrogenic} and it can be difficult to distinguish the effects of dielectric screening from direct electronic coupling.

To investigate electronic coupling at a QD/TMD interface, we use time-resolved second harmonic generation (SHG). Monolayer 2H TMD materials, such as the WS$_2$ used in this study, are highly nonlinear and lack inversion symmetry. This results in a large second order nonlinear susceptibility that depends on the number of TMD layers.\cite{MoS2_SHGImage_Heinz,MoS2_Imaging_PRB} Meanwhile, a film of randomly oriented CdSe quantum dots has no appreciable SHG response. Furthermore, SHG is sensitive to charge transfer at polarized interfaces, rendering it uniquely well-suited for studying interfacial dynamics.\cite{Tisdale2010_SciencePaper} 
\FloatBarrier
\begin{figure}[h!]
\begin{center}
\includegraphics[width=0.5\columnwidth]{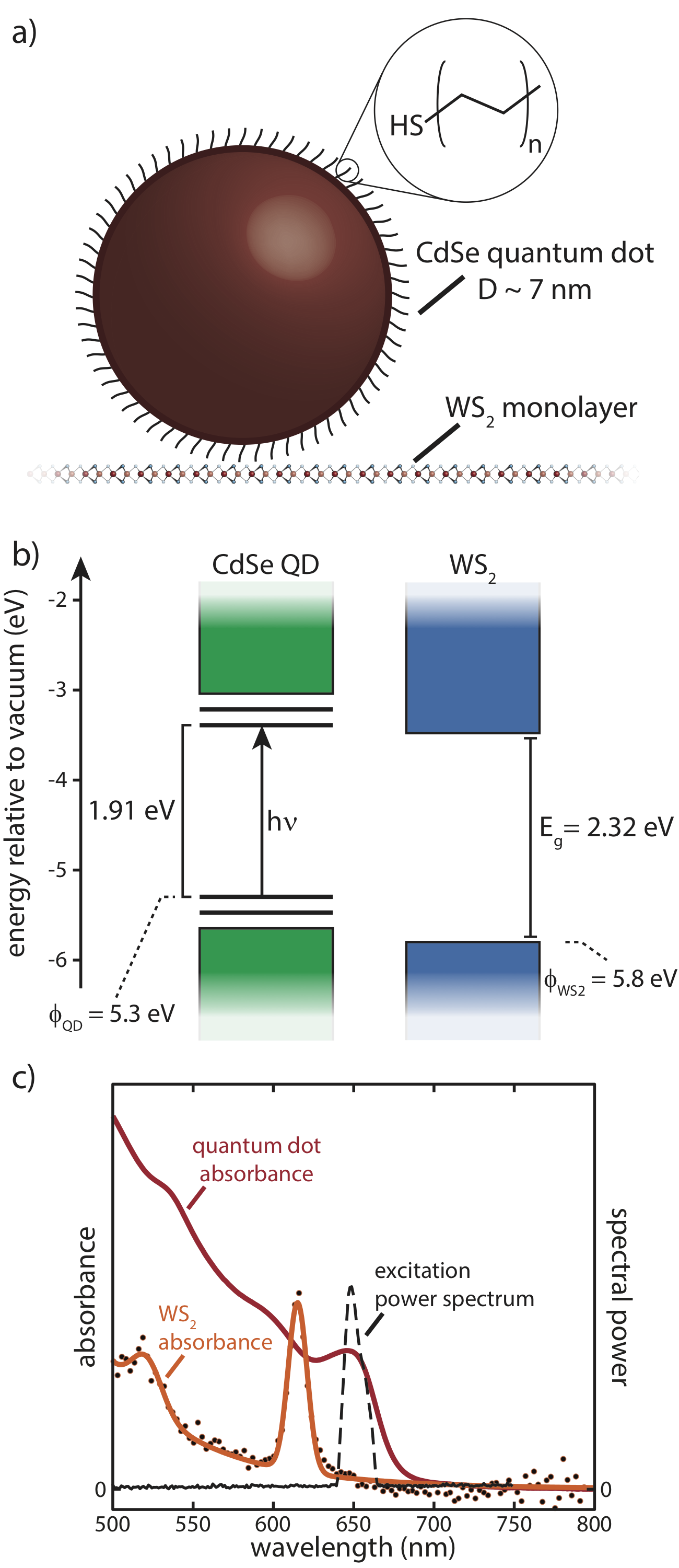}
\end{center}
\caption{\textit{Characterization of the quantum dot (QD)/WS$_2$ heterostructure.} (a) Schematic of the QD/WS$_2$ interface. QDs are coated with alkanethiol ligands of tunable length. (b) Estimated band alignment at the hybrid interface, based on published ionization potentials, exciton binding energies, and measured optical gaps. (c) CdSe QD (dark, red) and WS$_2$ (light, orange) absorption spectra. The pump pulse spectrum (black, dashed) selectively excites the quantum dots in the heterostructure.}
\label{fig1}
\end{figure}
\FloatBarrier

WS$_2$ monolayers were mechanically exfoliated from a bulk single crystal (2D Semiconductors) on a Si/SiO$_2$ substrate. CdSe QDs were synthesized by a previously reported method\cite{CdSe_synthetic_method} and suspended in toluene. The QD excitonic peak absorption occurred at $\lambda = 646$ nm (Fig.~\ref{fig1}c) corresponding to nanocrystals roughly 7 nm in diameter.\cite{jolene_sizing_curve} QDs were deposited onto the WS$_2$/SiO$_2$/Si substrate by spin coating and the native octadecylphosphonic acid surface ligands were subsequently exchanged for variable length alkanethiols in a layer-by-layer process.\cite{Mark_spincoating_method} In this method, $\sim 50$ $\mu$L of a 10 mg/mL QD suspension in toluene was spin-coated onto the WS$_2$/SiO$_2$/Si substrate at 1500 rpm for 30s. The surface was then covered with a 0.1 M solution of the desired alkanethiol (ethanethiol, 1-butanethiol, 1-octanethiol, or 1-dodecanethiol) in acetonitrile and allowed to soak for 30 s. The exchange solution was then spun off. Lastly, the sample was covered with pure acetonitrile and spun at 1500 rpm for 30 s to wash away free ligand. This process was repeated five times to form the nanocrystal film. The resulting interface is illustrated in Fig.~\ref{fig1}a.

The estimated band edge alignment at the QD/WS$_2$ interface is shown in Fig.~\ref{fig1}b. The valence band maxima for the CdSe cores and WS$_2$ relative to vacuum (the ionization potential, $\phi$) were taken from literature. Jasieniak \textit{et al.} reported size-dependent CdSe QD valence band maxima energies, which the measured using photoelectron spectroscopy.\cite{QD_BandEnergies_ACSNano2011} Keyshar \textit{et al.} reported the work functions of monolayer TMDs on SiO$_2$ measured using photoelectron microscopy,\cite{TMD_workFunction_Exp} reporting values for WS$_2$ in good agreement with electronic structure calculations.\cite{TMDC_BandEnergies_APL2013} Meanwhile, the energy of the excitonic transitions in WS$_2$ and our CdSe QDs are easily obtained from absorption spectroscopy. The WS$_2$ band gap was inferred by adding the reported exciton binding energy\cite{WS2_nonhydrogenic} to the exciton transition energy yielding a gap, E$_\textrm{g}$, in good agreement with scanning tunneling spectroscopy experiments.\cite{MoS2_WS2_bandAlignment_STM_Heinz_Nanolett_2016} The QD layer and WS$_2$ monolayer form a type-II heterojunction, which favors electron transfer from photoexcited QDs to WS$_2$. The QD (solution and film, see Supporting Information) and WS$_2$ absorbance are shown in Fig.~\ref{fig1}c; the QD absorbance extends to significantly lower energy than the WS$_2$ monolayer, enabling selective photoexcitation of the QD in the heterostructure.

\begin{figure}[h!]
	\begin{center}
		\includegraphics[width=1\columnwidth]{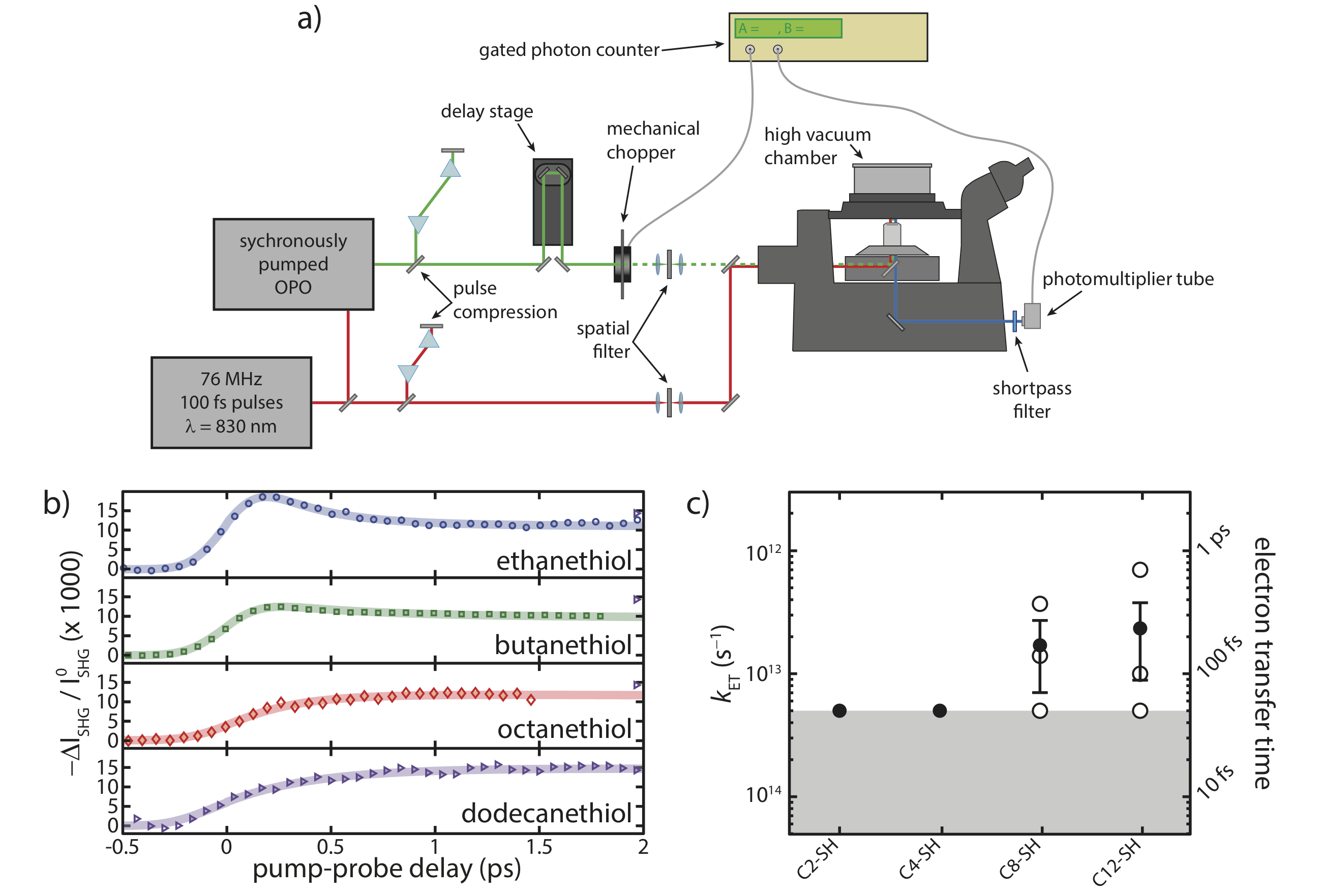}
	\end{center}
	\caption{\textit{Time-resolved second harmonic generation (SHG).} (a) Schematic of the time-resolved SHG microscopy instrument. (b) Transient SHG response of different samples. Data are shown for quantum dots capped with ethanethiol (blue, circles), butanethiol (green, squares), octanethiol (red, diamonds), and dodecanethiol (violet, triangles). Symbols represent the experimental data, while transparent lines represent fits to a simple kinetic model described by eqns.~(\ref{eqn1}-\ref{eqn2}). (c) Inverse electron transfer rate constants determined by fitting the data to our kinetic model. Values measured at different locations on the spatially heterogeneous sample are indicated by open circles, while the average values are denoted by filled circles. The gray line at the bottom of the figure represents the limitation of our instrument resolution.}

	\label{fig2}
\end{figure}

Our time-resolved SHG microscopy instrument is illustrated in Fig.~\ref{fig2}a. A 76 MHz Ti:sapphire oscillator (Coherent, Mira-HP) generates 100 fs, $\lambda = 830$ nm pulses, which synchronously pump an optical parametric oscillator (OPO, Coherent, Mira-OPO). The OPO generates 100 fs pulses of tunable wavelength ($\lambda = 550$ to 700 nm), which we use to photoexcite the sample. A 0.15 mJ/cm$^2$ pulse at $\lambda=650$ nm excites the QDs, and after a variable waiting period, SHG response of the sample is probed by a $\lambda = 830$ nm, 2~mJ/cm$^2$ pulse. Some fraction of the probe pulse is converted to the second harmonic, which is then transmitted through a short pass filter and detected by a photomultiplier tube (PMT). The pump-induced change in the SHG response is found by mechanically chopping the pump and using a gated counter to count SHG photons in the presence and absence of the pump pulse. Prior to entering the microscope, the pump and probe lines are spatially filtered and collimated to form diffraction-limited Gaussian spots at the sample (spot size $\sim 800$ nm in diameter). The sample was kept under vacuum in a sealed vacuum chamber, situated on top of the microscope stage, to prevent photoinduced damage in air. The smallest transfer times resolvable by SHG are limited the durations of the pump and probe pulses, which are measured by collecting the pump-probe sum frequency generation cross correlation (see Supporting Information).

The rate of electron transfer, which reflects the degree of electronic coupling in the hybrid system, is captured in early-time dynamics of the transient SHG response. We present the early time transient SHG signal, plotted as the change in the SHG intensity as a fraction of the intensity in the absence of the pump, in Fig.~\ref{fig2}b. Experimental data are plotted as open symbols while fits to a simple kinetic model described later are plotted as solid lines. When bare WS$_2$ is excited with the $\lambda=650$~nm pump, no transient SHG response is observed (see Supporting Information). In contrast, excitation of the QD/WS$_2$ heterostructure results in a sub-picosecond change in the SHG intensity. QDs capped with ethanethiol show the fastest rise time in the SHG signal, reflecting the fastest electron transfer rate. Longer alkanethiol capping ligands result in monotonically decreasing electron transfer rates as reflected in the elongated rise times. 

The ethanethiol sample additionally shows a fast recovery of the SHG signal, likely reflecting direct electron-hole recombination across the interface after the initial electron transfer event, facilitated by particularly strong electronic coupling with this short ligand. The fast recovery feature is less prominent in butanethiol coated dots and disappears entirely for octanethiol and dodecanethiol coated QDs. Fig.~\ref{fig2}c shows the extracted electron transfer rate constants $k_\mathrm{ET} = \tau_\mathrm{ET}^{-1}$ for QDs capped with each ligand. The open circles represent values extracted from experiments performed at different locations on the sample. Filled circles represent the average time constant value (error bars represent the standard error). There is considerable variation in the electron transfer time constant at different sample locations, reflecting heterogeneity in the QD/WS$_2$ interface -- presumably due to inconsistency in the degree of ligand coverage and orientation of the faceted QDs on the WS$_2$ surface. However, on average, we observe a monotonic increase in the electron transfer time constant as the capping ligand length is increased.

The observed electron transfer rate constants range from $1.4$ to $>100$ ps$^{-1}$, which is surprisingly fast for electron transfer from a CdSe QD. For comparison, optimal electron hopping rates between CdSe cores separated by atomically thin inorganic linkers are calculated to be 1 ps$^{-1}$ even when the interdot geometry is optimized for transfer.\cite{ET_hoppingRates_Calc} Effective hopping rate constants in the most conductive CdSe QD solids are at most $\sim 4$ ps$^{-1}$.\cite{ETHopping_Guyot_perspective} One possible explanation for the transient SHG signal is ultrafast hole trapping at the QD surface.\cite{hole_trapping} Though this process is known to occur in CdSe QDs, the transient SHG signal dependence on capping ligand length makes such an interpretation less likely.

\begin{figure}[h!]
	\begin{center}
		\includegraphics[width=1\columnwidth]{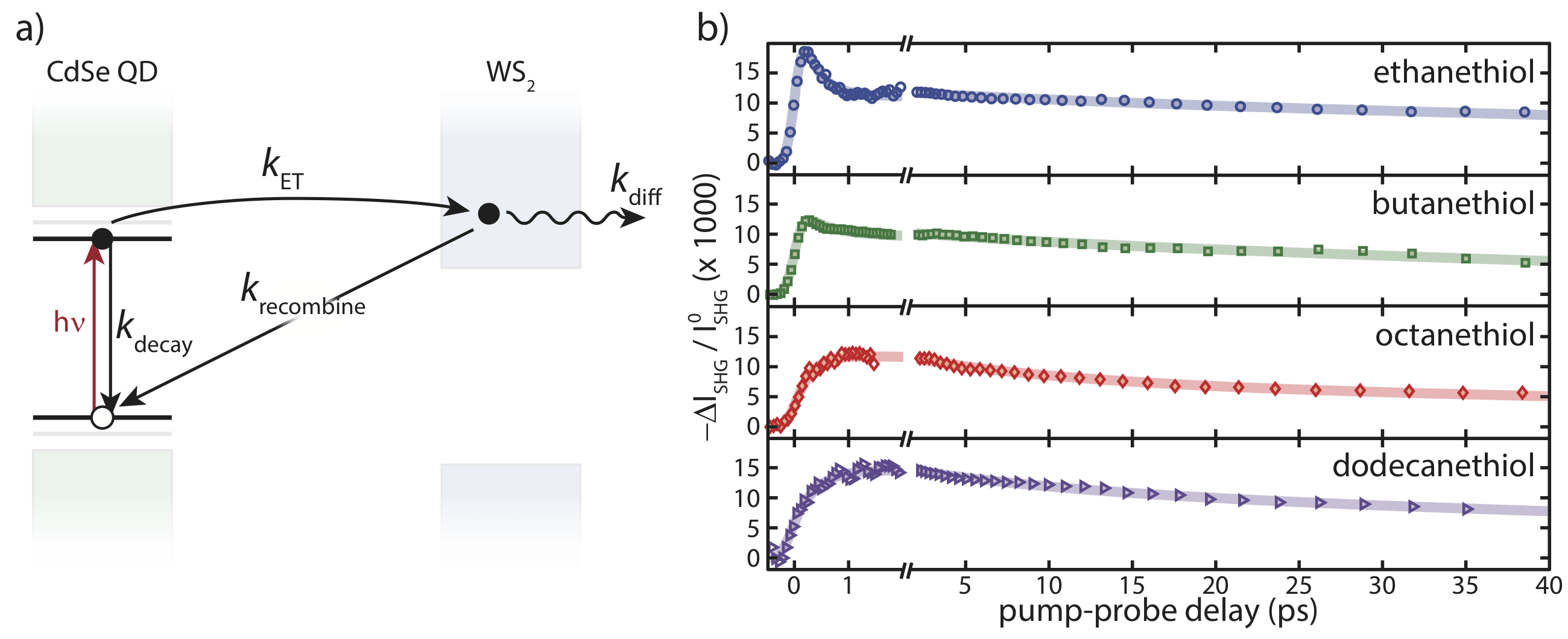}
	\end{center}
	\caption{\textit{Recovery of the transient SHG response.} (a) Following photoexcitation, an electron in the QD excited state  can return to the QD ground state or transfer to the acceptor WS$_2$. From there, the charge can recombine directly to the QD or diffuse within the WS$_2$. (b) Representative SHG signal recovery dynamics are shown for WS$_2$ covered with ethanethiol- (blue, cirlces), butanethiol- (green, squares), octanethiol- (red, diamonds), and dodecanethiol- (violet, triangles) capped QDs. Across multiple samples, long time transient SHG traces show similar relaxation dynamics. Dynamics are modelled by eqns.~(\ref{eqn1}-\ref{eqn2}).}
	\label{fig2p1}
\end{figure}

Fig.~\ref{fig2p1}a illustrates several possible pathways available to a photoexcited charge in this experiment. The excited electron can relax back to the QD ground state before it has a chance to transfer $\left(k_\mathrm{decay}\right)$. Alternatively, it can transfer to the neighboring WS$_2$ conduction band with the electron transfer rate constant $k_\mathrm{ET}$. Once transferred, the electron can recombine directly to the QD ground state with rate constant $k_\mathrm{recombine}$ or diffuse in the WS$_2$ plane, leading to an effective relaxation rate constant $k_\mathrm{diff}$. SHG data were fit to this kinetic model using the coupled differential equations,
\begin{align}
	\frac{\mathrm{d}\left[\mathrm{QD}^*\right]}{\mathrm{d}t} &= -(k_\mathrm{decay}+k_\mathrm{ET})\left[\mathrm{QD}^*\right] \label{eqn1}\\
	\frac{\mathrm{d}\left[\mathrm{WS}_2^*\right]}{\mathrm{d}t} &= k_\mathrm{ET}\left[\mathrm{QD}^*\right]-(k_\textrm{diff}+k_\mathrm{recombine})\left[\mathrm{WS}_2^*\right].\label{eqn2}
\end{align}
The time-dependent area densities of excited charges on QDs, $\left[\mathrm{QD}^*\right](t)$, and WS$_2$, $\left[\mathrm{WS}_2^*\right](t)$, were initialized with initial condition $\left[\mathrm{QD}^*\right](0)=N_0$ and $\left[\mathrm{WS}_2^*\right](0)=0$, reflecting QD-selective excitation. Traces were fit to this kinetic model to extract key rates such as $k_\mathrm{ET}$ (Fig.~\ref{fig2}b). For direct comparison to experimental data, the modeled kinetics were convolved with the pump-probe sum frequency generation cross correlation.

Fig.~\ref{fig2p1}b shows the SHG signal recovery dynamics for samples coated with the four capping ligands.  The decaying portions of the traces are fit to biexponentials, which are plotted as thick, transparent lines. The long time dynamics reflect a complicated set of processes that occur following electron transfer, including lateral diffusion within the WS$_2$ plane and subsequent recombination with a hole. The long-time relaxation rate constant, $k_\mathrm{diff}$, varied somewhat by sample location but did not show a clear dependence on QD ligand.

\begin{figure}[h!]
	\begin{center}
		\includegraphics[width=1\columnwidth]{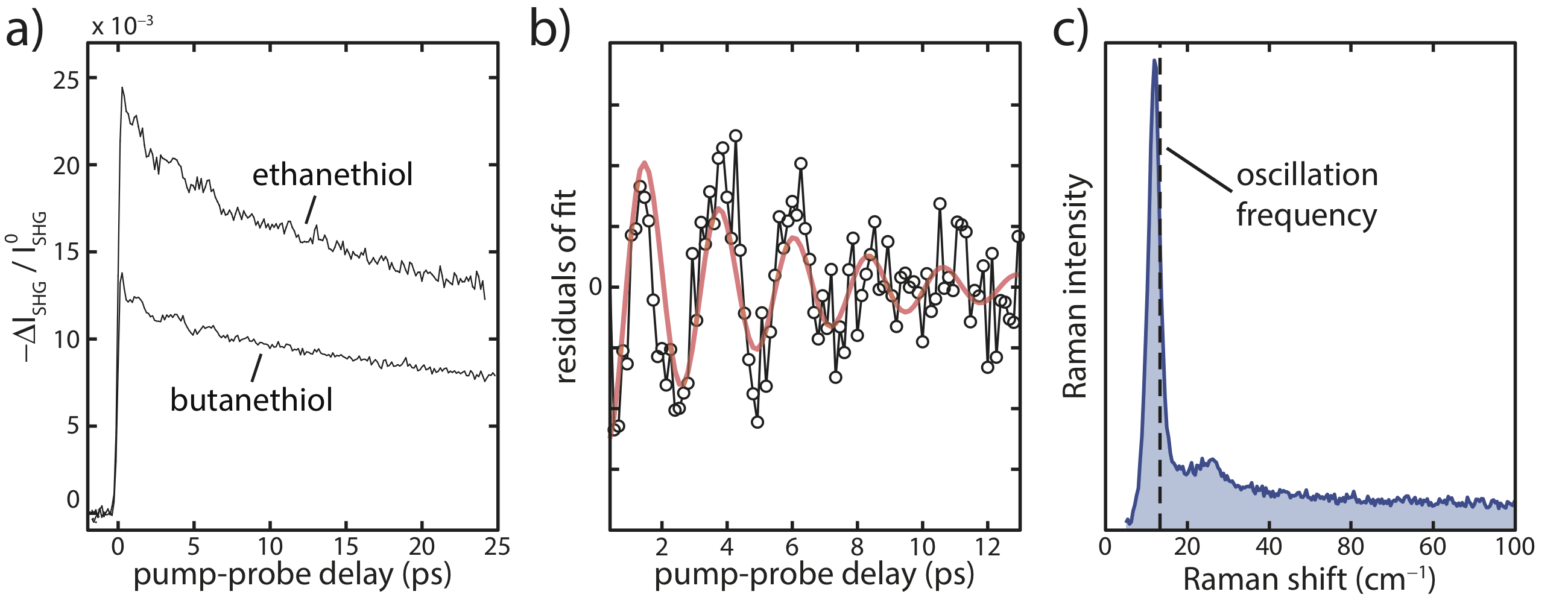}
	\end{center}
	\caption{\textit{Coherent phonon oscillations.} (a) Transient SHG response for ethanethiol- and butanethiol-capped QD samples on WS$_2$, exhibiting coherent oscillation in the signal intensity. (b) Residuals of a bi-exponential fit to the ethanethiol data in (a) are plotted in black, and fitted to a damped sine wave, plotted in red, with a frequency of $\sim 14$ cm$^{-1}$. (c) The fit extracted an oscillation frequency $\sim 14$ cm$^{-1}$. (c) Non-resonant Raman spectrum of the same QDs, plotted in blue. The radially symmetric QD breathing mode scatters inelastically at a frequency matching the picosecond oscillations in the transient SHG data.}
	\label{fig3}
\end{figure}

QDs capped with ethanethiol and butanethiol exhibited the fastest electron transfer dynamics. When transient SHG signals from these samples were collected for longer integration times (to improve signal-to-noise ratio), coherent oscillations in the transient SHG data became apparent (Fig.~\ref{fig3}a). These oscillations were not observed in WS$_2$-only or QD-only samples. Notably, oscillations were also not observed in octanethiol and dodecanethiol samples, indicating that strong electronic coupling -- rather than QD photoexcitation alone -- is required for their appearance in the transient SHG signal. The oscillatory component of the transient SHG signal was isolated by fitting a biexponential decay to the data and plotting the residuals of the fit. Residuals from the sample with butanethiol-capped QDs are plotted in Fig.~\ref{fig3}b (black markers) along with a damped sine wave (red line). The fitted frequency of this oscillatory component is $\Omega = 14.0$ cm$^{-1}$ $\pm$ 0.6 cm$^{-1}$.

QDs exhibit quantized low-frequency acoustic vibrations analogous to those of an elastic sphere.\cite{Jolene_TempDep_LowFreqRam,Liza_ElasticContinuum_Raman,size_dep_vib_CdSeQD,Size_DependentDynamics_Sphere}  The lowest energy collective vibration corresponds to a radially symmetric breathing mode. This mode is Raman active, and appears in the low-frequency Raman spectrum of the dots used in this system (collected with a previously reported\cite{jolene_sizing_curve} experimental apparatus). The Raman spectrum scattered by the 7 nm diameter QDs is plotted in blue in Fig.~\ref{fig3}c. The 14 cm$^{-1}$ Raman breathing mode matches the frequency of the picosecond oscillations in the transient SHG traces. The fitted damping rate, $\Gamma \sim 0.2$ ps$^{-1}$, corresponds to a $\sim 6$ cm$^{-1}$ linewidth, which is similar to the acoustic phonon linewidth in the Raman spectrum ($\sim$4 cm$^{-1}$) and consistent with other time-domain measurements performed on CdSe QDs in solution.\cite{CoherentPhonons_CdSe_PatKam_PRB}

It is significant that a coherent vibration of the CdSe QD is observed, even though it is the WS$_2$ layer that contributes the SHG signal we measure (CdSe films deposited on glass substrates generated no detectable SHG signal in our instrument). There are two possible explanations for this: 1) strong dielectric coupling, or 2) strong electronic-vibrational coupling at the QD-TMD interface. In the first case, coherent expansion and compression of QDs situated on top of the WS$_2$ surface periodically modulates the dielectric environment surrounding the 2D material, which in turn modulates the nonlinear susceptibility. However, coherent acoustic phonons in QDs can be excited directly upon photoexcitation,\cite{Size_DependentDynamics_Sphere,CoherentPhonons_CdSe_PatKam_PRB} whereas we only observe coherent oscillations in the case of ultrafast charge transfer. 

That coherent oscillations are only observed for the two shortest molecular ligands suggests that explanation $\#$2 -- strong electronic-vibrational coupling at the CdSe QD/WS$_2$ interface -- is the likely mechanism. In this mechanism, the adiabatic transition state for the donor-acceptor electron transfer process is mixed due to strong electronic coupling.\cite{coherent_CT_org_photovo_blend,Scholes_Coherence_Lett_2018,ArtificialAtoms_Will_PNAS2011} Acoustic vibration of the donor QD coherently modulates the electronic coupling strength, creating a vibronic signature in the time-domain second-order nonlinear optical response.

In conclusion, we engineered a QD/TMD system to facilitate fast charge transfer by placing CdSe QD cores directly on a WS$_2$ monolayer separated by short alkanethiol ligands. We adjusted the energetics of the system to exclude all processes except charge transfer and probed the charge transfer dynamics using time-resolved SHG. We were able to tune the rate of charge transfer by changing the QD capping ligand, adjusting the degree of electronic coupling at the QD/WS$_2$ interface. In the case of ethanethiol and butanethiol, the ultrafast transfer process coherently excited a QD acoustic phonon, modulated the SHG response in the time domain. Charge transfer is a fundamental process underlying 0D/2D hybrid optoelectronic devices and this work probes that process spectroscopically at a model interface. The work also demonstrates the advantages of using SHG to probe dynamics at TMD interfaces, which can be difficult to study with linear spectroscopy.

\section{Acknowledgments}
Fabrication of QD/TMD heterostructures was supported by the Samsung Global Research Opportunities (GRO) program. Development of the transient SHG microscope was supported by the U.S. Department of Energy, Office of Basic Energy Sciences, Division of Chemical Sciences, Geosciences \& Biosciences under Award Number DE-SC0010538.

\FloatBarrier
\bibliography{ETBib}
\FloatBarrier

\newpage
\section{Supporting Information}
\subsection{Transient SHG Response of a Bare WS$_2$ Monolayer}
\FloatBarrier
\begin{figure}[h!]
	\begin{center}
		\includegraphics[width=1\columnwidth]{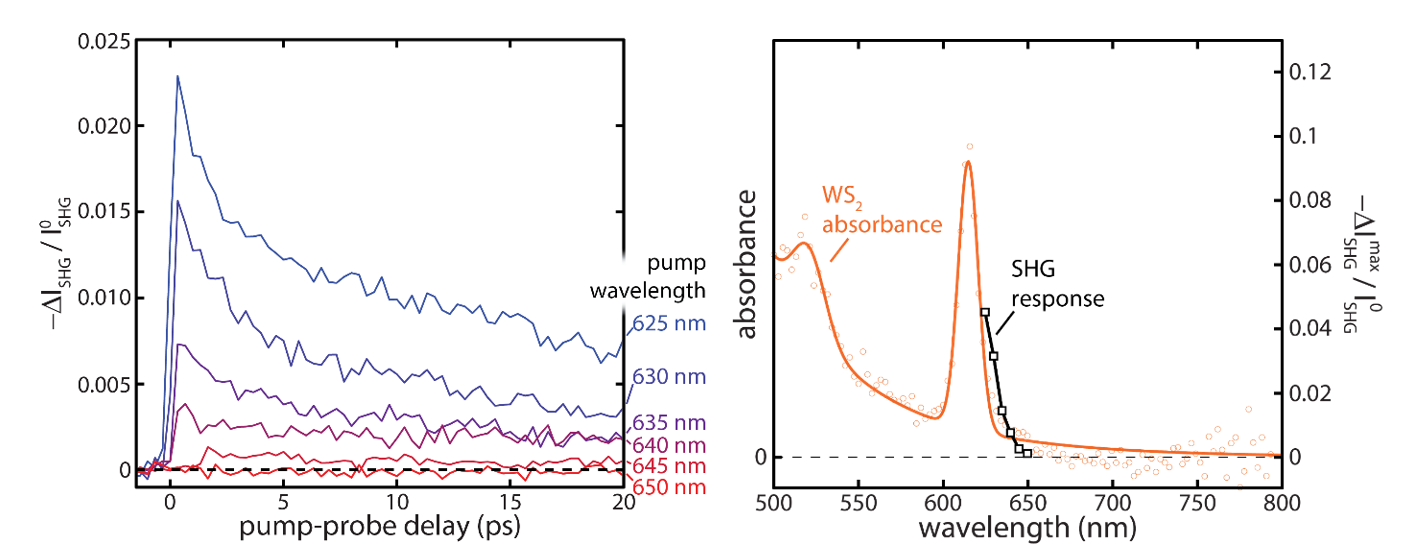}
	\end{center}
	\caption{\textit{left:} Transient SHG response of bare WS$_2$ monolayer. The colored traces represent experiments performed with different excitation wavelengths as indicated. The fluence is kept constant. \textit{Right:} The maximum transient SHG response for a given excitation wavelength (black line) is plotted on the same graph as the bare WS$_2$ absorbance spectrum (orange line). The magnitude of the transient SHG response closely follows the WS$_2$ absorption spectrum.}
\end{figure}

To show that the transient SHG signals observed in the main text arose from absorption by the CdSe dots -- and not by direct photoexcitation of the WS$_2$ layer -- we  performed time-resolved SHG microscopy on a bare WS$_2$ monolayer. Though the monolayer shows appreciable transient response when excited by $\lambda =$ 625, 630, 635, 640, and 645 nm pump pulses, it shows no discernible response when excited with the excitation spectrum used in the main text ($\lambda =$ 650 nm). The transient signals for QDs-on-WS$_2$ presented in the main text reached maxima of roughly $-\frac{\Delta I_\mathrm{SHG}}{I_\mathrm{SHG}^0}\approx 0.01~-~0.02$ at 650 nm excitation wavelength, which is more than an order of magnitude larger than the transient SHG response of the bare WS$_2$ at the same excitation wavelength. 
\FloatBarrier

\newpage
\subsection{QD Film Absorbance}
\FloatBarrier
\begin{figure}[h!]
	\begin{center}
		\includegraphics[width=.65\columnwidth]{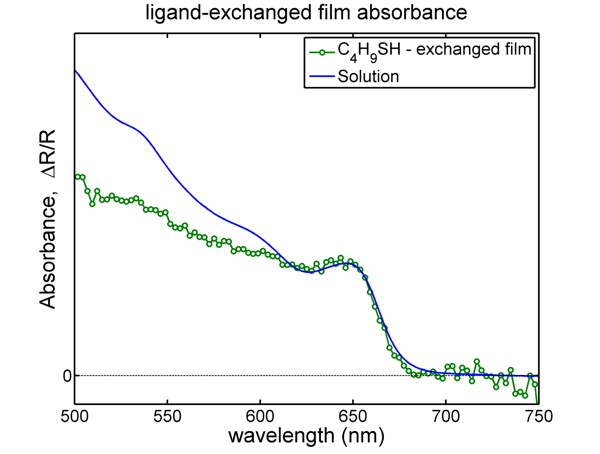}
	\end{center}
	\caption{The QD solution absorbance spectrum presented in the main text is plotted (blue line) with the QD film reflectivity spectrum ($\Delta$R/R) after butane-thiol ligand exchange. The ground state absorption spectrum is not changed near the excitation resonance following ligand exchange.}
\end{figure}
\FloatBarrier

\newpage
\subsection{Pump-Probe SFG Cross Correlation}
\FloatBarrier
\begin{figure}[h!]
	\begin{center}
		\includegraphics[width=1\columnwidth]{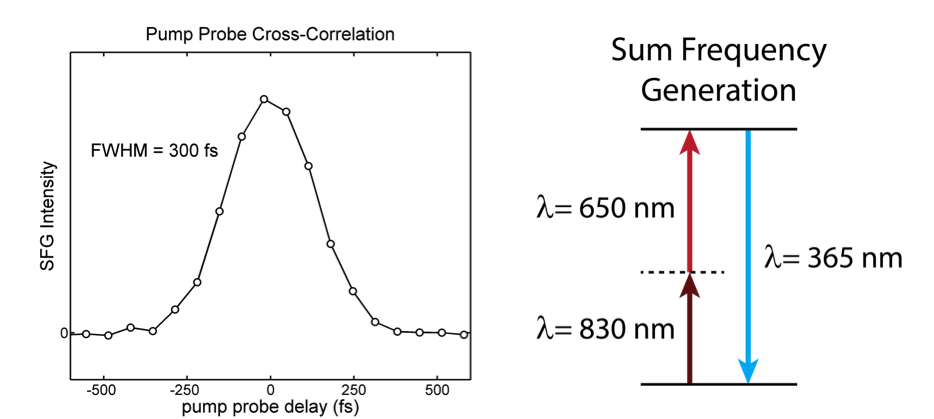}
	\end{center}
	\caption{\textit{Time resolution of the SHG microscope.} Sum frequency generation (SFG) was performed by mixing the pump and probe pulses in monolayer WS$_2$. The SFG signal was filtered by a short pass filter to remove the pump, probe, and the probe second harmonic, and detected with the same PMT used in time-resolved SHG experiments. The SFG cross correlation represents the convolution of the pump and probe intensity profiles.}
\end{figure}

\end{document}